\documentclass[two column, prA]{revtex4}%
\usepackage{amsmath}
\usepackage{graphicx}
\usepackage{amsfonts}
\usepackage{amssymb}%
\providecommand{\U}[1]{\protect\rule{.1in}{.1in}}

\begin{document}
\title{Photon correlations in positron annihilation }
\author{Isabelle Gauthier and Margaret Hawton}
\email{margaret.hawton@lakeheadu.ca}
\affiliation{Department of Physics, Lakehead University, Thunder Bay, ON, Canada, P7B 5E1}

\begin{abstract}
The two-photon positron annihilation density matrix is found to separate into
a diagonal center of energy factor implying maximally entangled momenta, and a
relative factor describing decay. For unknown positron injection time, the
distribution of the difference in photon arrival times is a double exponential
at the para-Ps decay rate, consistent with experiment (V. D. Irby, Meas. Sci.
Technol. \textbf{15}, 1799 (2004)).

\end{abstract}
\maketitle

\section{Introduction}

When an electron and a positron with opposite spin annihilate, two correlated
photons with total energy $2\times.511\ MeV$ are created. These annihilation
$\gamma$-rays cannot be manipulated using optical beam splitters and mirrors,
so interference experiments and applications in quantum information are not
practical. However, positron annihilation is important in medicine and
material science \cite{PASreview}. In medical imaging, coincident detection of
the annihilation photons is the basis for positron emission tomography (PET).
In material science positron annihilation spectroscopy (PAS) gives information
on electron density and the distribution of electron momenta.

Positrons are created by the decay of radioactive nuclei such as $^{22}Na$ or
$^{18}F$ imbedded in the sample of interest. For example, the $1.275\ MeV$
nuclear $\gamma$-ray emitted immediately following the positron emission from
$^{22}Na$ determines the time of positron injection. In positron lifetime
(PAL) measurements the arrival time difference between the nuclear photon and
one of the annihilation photons is measured. Positron annihilation in
condensed matter proceeds through bound states of positrons with electrons,
atoms, molecules and various defects \cite{PASreview}. The annihilating
positron and electron form a free or bound hydrogen-like positronium (Ps)
atom. In vacuum, singlet or para-Ps decays into two $\gamma$-rays with a
lifetime of $125\ ps$. In $\alpha-$SiO$_{2}$ the para-Ps lifetime is increased
to $156\ ps$ due to modification of the dielectric constant and electron mass
relative to vacuum \cite{Saito}.

Recently it has been suggested that measurement of the arrival time difference
between paired annihilation photons will improve signal to noise in medical
imaging applications, leading to time of flight (TOF) PET \cite{Lewellen}.
This is plausible because the most widely accepted viewpoint is that the
minimum quantum uncertainty in time is zero due to detection-induced nonlocal
collapse \cite{IrbyExpt}. Irby measured the time interval between detection of
the annihilation photons from a $^{22}Na$ source \ and obtained $123\pm22\ ps$
\cite{IrbyExpt}. This is a surprising result since, in his experiment, the
annihilation photons originate in a source a few $mm$ thick and a photon
travels almost $4\ cm$ in air in this time.

To explain these observations, Irby generalized the Einstein, Podolsky and
Rosen (EPR) \cite{EPR} example of position and momentum as elements of reality
to include time and energy dependence \cite{IrbyTheory}. Using entangled spins
as an illustration, he showed that restriction of one observable leads to
reduced nonlocality of its conjugate. He attributed his experimental results
to maximally restricted photon momenta, leading to elimination of nonlocality
in the conjugate position observables. However, a complete explanation
requires a theory of the $123\ ps$ wide distribution of time differences that
he observed. Here we give a quantitative explanation of his observations by
performing a detailed analysis of Ps decay.

\section{Theory}

This section is based on Sakurai's theory of positron annihilation
\cite{Sakurai}, summarized in Subsection A, transformed to relative and center
of energy coordinates in Subsection B, and modified to explicitly include
exponential decay in Subsection C. Natural units in which $\hbar=c=1$ are
used, the electron/positron mass is denoted as $m$, and the positron charge is
$e$. The dimensionless fine structure constant is then $\alpha=e^{2}%
/4\pi=1/137_{.}$ The subscript $+$ refers to the positron and $-$ to an
electron. We consider a relativistic expansion in powers of the Fermion
speeds, $\beta_{+}$ and $\beta_{-}$, denoted $\beta_{\pm}$ where, to first
order in $\beta_{\pm}$, the annihilation photons are counterpropagating. To
simplify the equations it is assumed that the photon pulses are well separated
from the positron source when they reach the detectors.%

\begin{figure}
[ptb]
\begin{center}
\includegraphics[
height=1.8836in,
width=3.0415in
]%
{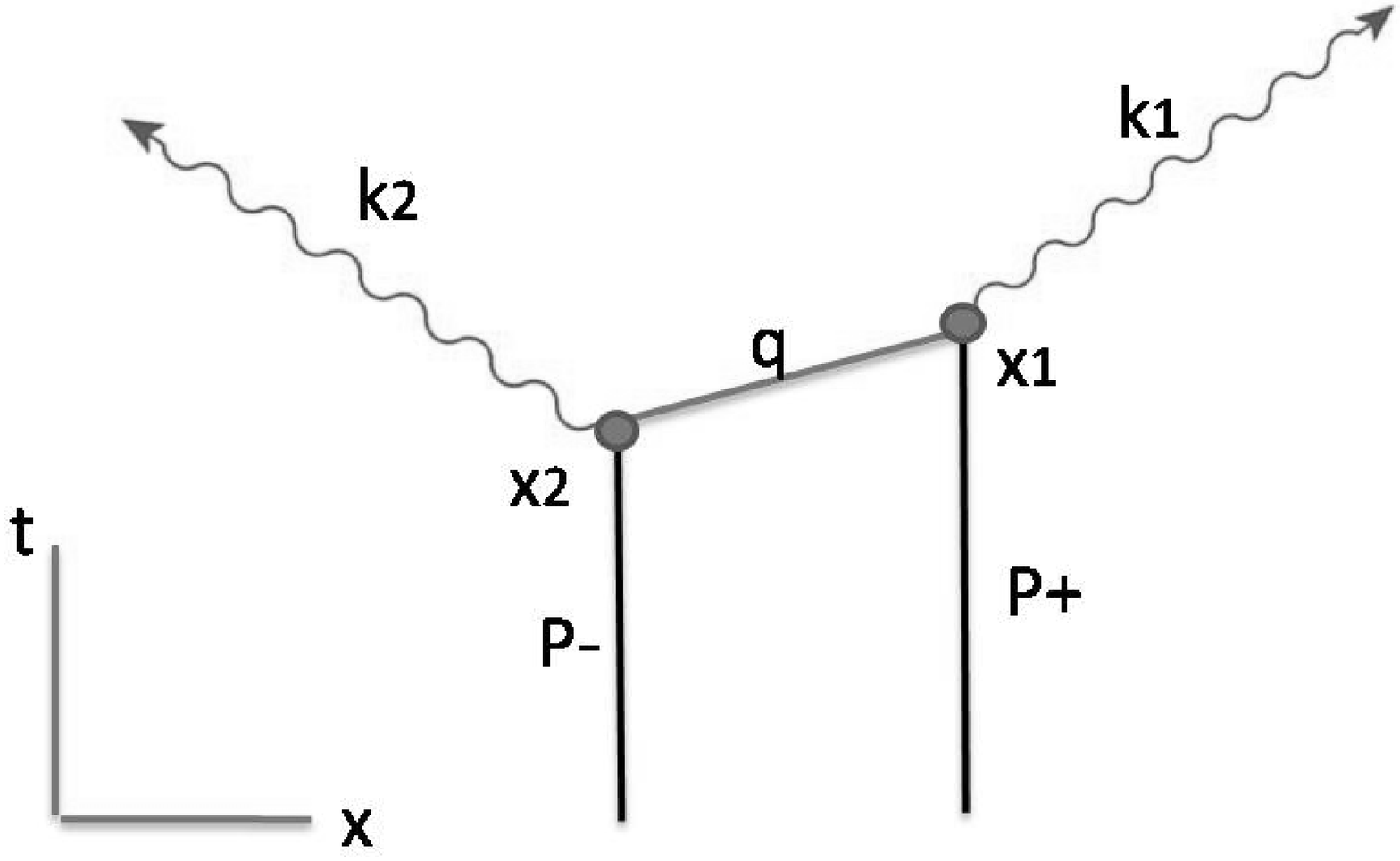}%
\caption{A two photon Feynman diagram. An electron, $p_{-},$ emits a photon,
$k_{2},$ while scattering to a virtual state, $q$. It then annihilates with a
positron, $p_{+}$, while creating a second photon, $k_{1}.$}%
\end{center}
\end{figure}

\subsection{Positron annihilation}

Position annihilation according to the Dirac equation is discussed by Sakurai.
He performs a perturbation expansion in powers of $e$ and finds that the first
nonzero term is of second order. The Feynman diagram of such a process is
sketched in Fig. 1: An electron with four-momentum $p_{-}=\left(
E_{-},\mathbf{p}_{-}\right)  $ is scattered to four-momentum $q=\left(
q_{0},\mathbf{q}\right)  $ at space-time point $x_{2}=\left(  t_{2}%
,\mathbf{x}_{2}\right)  $ while emitting a photon with four-momentum
$k_{2}=\left(  \omega_{2},\mathbf{k}_{2}\right)  .$ At $x_{1}$ this electron
annihilates with the positron and emits a photon with four-momentum $k_{1}.$
If instead the positron is scattered first, $q\leftrightarrow-q$ and the
photons are interchanged. Sakurai obtains a scattering cross section for
two-photon annihilation of $\pi r_{0}^{2}/\beta_{+}$ where $r_{0}=\alpha/m$.
The Bohr radius, $a_{0}=1/\left(  \alpha m\right)  $, is larger than $r_{0}$
by a factor $\alpha^{-2}$, so the volume of an atom appears infinite on the
length scale $r_{0}$ and the center of energy momentum is conserved, that is%
\begin{equation}
\mathbf{k}_{1}+\mathbf{k}_{2}=\mathbf{p}_{+}+\mathbf{p}_{-}. \label{MomCons}%
\end{equation}
Sakurai applies his scattering theory to Ps by setting the electron density
equal to $\left\vert \psi_{1s}\right\vert ^{2}=1/\left[  \pi\left(
2a_{0}\right)  ^{3}\right]  $ and obtains a decay rate%
\begin{equation}
\Gamma=\frac{1}{2}\alpha^{5}m, \label{Gamma}%
\end{equation}
equivalent to a lifetime $\Gamma^{-1}=125\ ps$.

In Sakurai's covariant formulation energy and momentum are conserved at the
vertices and the state $q$ describes a virtual particle for which the Fermion
dispersion relation is not imposed. However, since the final and initial
states describe real particles, the dispersion relations
\begin{align}
\omega_{j}  &  =\left\vert \mathbf{k}_{j}\right\vert ,\label{dispersion}\\
E_{\pm}  &  =\sqrt{m^{2}+\left\vert \mathbf{p}_{\pm}\right\vert ^{2}}\nonumber
\end{align}
must be satisfied. In the more usual noncovariant formulation of perturbation
theory, the dispersion relation is satisfied by the virtual Fermion but energy
is not conserved between $t_{1}$ and $t_{2}$. To zero order in $\beta_{\pm}$
the annihilation photon $k_{2}$ has energy $m$, so the excess energy of the
virtual state must be greater than $m$. Thus the intermediate state in Fig. 1
persists for less than $m^{-1}=1.3\times10^{-21}\ s,$ implying that two photon
annihilation is effectively instantaneous.

\subsection{Relative and center coordinates}

Here the center (of energy) and relative coordinates,%
\begin{align}
\mathbf{k}_{c}  &  =\mathbf{k}_{1}+\mathbf{k}_{2},\ \mathbf{k}_{r}=\frac{1}%
{2}\left(  \mathbf{k}_{1}-\mathbf{k}_{2}\right)  ,\label{coord}\\
\mathbf{p}_{c}  &  =\mathbf{p}_{+}+\mathbf{p}_{-},\ \ \mathbf{p}_{r}=\frac
{1}{2}\left(  \mathbf{p}_{+}-\mathbf{p}_{-}\right)  ,\ \nonumber\\
\mathbf{x}_{c}  &  =\frac{1}{2}\left(  \mathbf{x}_{1}+\mathbf{x}_{2}\right)
,\ \text{and }\mathbf{x}_{r}=\mathbf{x}_{1}-\mathbf{x}_{2},\nonumber
\end{align}
will be used. Since $\mathbf{k}_{1}\cdot\mathbf{x}_{1}+\mathbf{k}_{2}%
\cdot\mathbf{x}_{2}=\mathbf{k}_{c}\cdot\mathbf{x}_{c}+\mathbf{k}_{r}%
\cdot\mathbf{x}_{r}$ for the photons and $\mathbf{p}_{+}\cdot\mathbf{x}%
_{1}+\mathbf{p}_{-}\cdot\mathbf{x}_{2}=\mathbf{p}_{c}\cdot\mathbf{x}%
_{c}+\mathbf{p}_{r}\cdot\mathbf{x}_{r}$ for the Fermions, the exponent in a
Fourier transform is preserved by this transformation, and relative momentum
and position are conjugate observables, as are center momentum and position.

For counterpropagating photons the magnitudes of $\mathbf{k}_{1}$ and
$\mathbf{k}_{2}$ should be added (subtracted) to obtain the magnitude of the
relative (center) wave vector so that, according to (\ref{dispersion}) and
(\ref{coord}),%
\begin{align}
\omega &  \equiv\omega_{1}+\omega_{2}=2\left\vert \mathbf{k}_{r}\right\vert
,\label{mag_k}\\
\Delta\omega &  \equiv\omega_{1}-\omega_{2}=\left\vert \mathbf{k}%
_{c}\right\vert .\nonumber
\end{align}
To second order in $\beta_{\pm}$ the Ps total energy is
\begin{equation}
E=2m+p_{c}^{2}/4m. \label{ECons}%
\end{equation}
For a positron created at time $t_{0}$ contributions with different $p_{c}$
rapidly get out of phase due to the factor $\exp\left[  -ip_{c}^{2}\left(
t-t_{0}\right)  /4m\right]  $, leading to a density matrix that is diagonal in
center of energy momentum. The relative dynamics, described by $\mathbf{k}%
_{r}$, is decoupled from the center motion, described by $\mathbf{k}_{c}$. In
relative and center coordinates conservation of momentum, (\ref{MomCons}),
becomes%
\begin{equation}
\mathbf{k}_{c}=\mathbf{p}_{c}. \label{kCons}%
\end{equation}
Since $\mathbf{p}_{c}$ has a definite value, the momenta of the annihilation
photons are maximally restricted according $\mathbf{k}_{2}=\mathbf{p}%
_{c}-\mathbf{k}_{1}$ as observed by Irby.

\subsection{Dynamics}

Sakurai calculates the Ps decay rate so, implicitly, $\omega$ isn't exactly
equal to $E,$ but has a linewidth, $\Gamma$. Decay as a function of $t$ will
be considered in this subsection.

A pure state will be written as a linear combination of a Ps atom in the $1s$
state with definite center of mass momentum $\mathbf{p}_{c}$, and the two
annihilation photons described by their relative and center momenta. If a
positron is injected at time $t_{0}$ the Schr\"{o}dinger picture (SP) state
vector is then%
\begin{equation}
\left\vert \Psi_{\mathbf{k}_{c}}\right\rangle =c_{1s}\left(  t\right)
\left\vert 1s,\mathbf{k}_{c}\right\rangle +\sum_{\mathbf{k}_{r}}%
c_{\mathbf{k}_{r}}\left(  t\right)  \left\vert \mathbf{k}_{r},\mathbf{k}%
_{c}\right\rangle . \label{State}%
\end{equation}
for $t>t_{0}$ and $\left\vert \Psi_{\mathbf{k}_{c}}\right\rangle =0$ for
$t<t_{0}$. We will take the volume, $V$, to be finite so that the momenta are
discrete. To second order in $e$ the dynamical equations describing the
relative motion for $t>t_{0}$ are \cite{SakuraiPage184}%
\begin{align}
\overset{\cdot}{c_{1s}}\left(  t\right)   &  =-iEc_{1s}\left(  t\right)
-i\sum_{\mathbf{k}_{r}}\ U_{r}^{(2)}c_{\mathbf{k}_{r}}\left(  t\right)
,\label{c_dot}\\
\overset{\cdot}{c}_{\mathbf{k}_{r}}\left(  t\right)   &  =-i\omega
c_{\mathbf{k}_{r}}\left(  t\right)  -iU_{r}^{(2)}\ c_{1s}\left(  t\right)
\nonumber
\end{align}
where the dot denotes differentiation with respect to $t$ and $\overset{\cdot
}{U}_{fi}^{(2)}=U_{r}^{(2)}\delta^{3}\left(  \mathbf{k}_{c}-\mathbf{p}%
_{c}\right)  $ is the time derivative of the transition matrix element from Ps
to the two-photon state. Eqs. (\ref{c_dot}) describe Weisskopf-Wigner
spontaneous emission that is exponential in time and Lorentzian in frequency.
A system of equations of the form (\ref{c_dot}) are solved in the interaction
picture in \cite{Scully}. For $t-t_{0}\gg\Gamma^{-1}$ decay is essentially
complete so that the photon pulse is well separated from the source and
\cite{ScullyEq18} gives
\begin{equation}
c_{\mathbf{k}_{r}}\left(  t\right)  =AU_{r}^{(2)}\frac{\exp\left[
-i\omega\left(  t-t_{0}\right)  \right]  }{\omega-E+i\Gamma} \label{c_w}%
\end{equation}
in the SP with $\omega=2\left\vert \mathbf{k}_{r}\right\vert $ and $E=2m$ to
first order in $\beta_{\pm}$. The factor $A$ is a constant and (\ref{c_w}) can
be normalized using the integral $I1$ in Appendix A with the result%
\begin{equation}
c_{\mathbf{k}_{r}}\left(  t\right)  =\sqrt{\frac{8\pi\Gamma}{VE^{2}}}%
\frac{\exp\left[  -i\omega\left(  t-t_{0}\right)  \right]  }{\omega-E+i\Gamma
}. \label{All_c}%
\end{equation}

A pure state vector is of the form%
\begin{equation}
\left\vert \Psi_{\mathbf{k}_{c}}\right\rangle =\Theta\left(  \tau_{1}%
-t_{0}\right)  \Theta\left(  \tau_{2}-t_{0}\right)  \left\vert \mathbf{k}%
_{c}\right\rangle \otimes\left\vert \Psi_{r}\right\rangle \label{AllState}%
\end{equation}%
\begin{equation}
\tau_{j}\equiv t-\left\vert \mathbf{x}_{j}\right\vert \label{tau}%
\end{equation}
where $\mathbf{x}_{j}$ is the position of the $j^{th}$ photon, $\tau_{j}$ is
its emission time, the $\Theta$-functions ensure that no photons exist before
the positron is injected, and%
\begin{equation}
\left\vert \Psi_{r}\right\rangle =\sqrt{\frac{8\pi\Gamma}{VE^{2}}}%
\sum_{\mathbf{k}_{r}}\frac{\exp\left[  -i\omega\left(  t-t_{0}\right)
\right]  }{\omega-E+i\Gamma}\left\vert \mathbf{k}_{r}\right\rangle
\label{Relative}%
\end{equation}
describes the relative dynamics.

The space-time wave function is $\psi\left(  \mathbf{x}_{r},t\right)
=\left\langle \mathbf{x}_{r}|\Psi_{r}\right\rangle $ such that
\begin{equation}
\left\vert \Psi_{r}\right\rangle =\int d^{3}x_{r}\psi\left(  \mathbf{x}%
_{r},t\right)  \left\vert \mathbf{x}_{r}\right\rangle \label{rState}%
\end{equation}
with%
\begin{align}
\psi\left(  \mathbf{x}_{r},t\right)   &  =\sqrt{\frac{4\Gamma}{E^{2}}}%
\sum_{\mathbf{k}_{r}}\frac{\exp\left(  i\omega t_{0}\right)  }{\omega
-E+i\Gamma}\label{Relative_x}\\
&  \times\frac{\exp\left(  i\mathbf{k}_{r}\cdot\mathbf{x}_{r}-i\omega
t\right)  }{\left(  2\pi\right)  ^{3/2}}.\nonumber
\end{align}
Strictly speaking, the $\mathbf{k}_{r}$-amplitudes should be weighted as in a
$1s$ state, but $\Gamma\ll a_{0}^{-1}$, so this can be ignored. Substitution
of $\mathbf{k}=\mathbf{k}_{r}$, $\mathbf{r}=\mathbf{x}_{r}$ and $t=t-t_{0}$ in
integral $I2$ of in Appendix B gives
\begin{align}
\psi\left(  \left\vert \mathbf{x}_{r}\right\vert ,t\right)   &  =\sqrt
{\frac{\Gamma}{4\pi}}\frac{1}{\left\vert \mathbf{x}_{r}\right\vert
}\label{Psi_rel}\\
&  \times\exp\left[  -\left(  iE+\Gamma\right)  \left(  t-t_{0}-\frac{1}%
{2}\left\vert \mathbf{x}_{r}\right\vert \right)  \right] \nonumber
\end{align}
\ where a similar term involving $t-t_{0}+\frac{1}{2}\left\vert \mathbf{x}%
_{r}\right\vert $\ has been neglected. This wave function is normalized if it
is assumed that the photon pulse has propagated far enough so that
$\exp\left[  -\Gamma\left(  t-t_{0}\right)  \right]  \ll1$.

For a measurement described by the operator $\widehat{O}$, the expected value
is
\begin{equation}
\left\langle \widehat{O}\right\rangle =\sum_{\mathbf{k}_{c}}p_{\mathbf{k}_{c}%
}\left\langle \Psi_{\mathbf{k}_{c}}\left\vert \widehat{O}\right\vert
\Psi_{\mathbf{k}_{c}}\right\rangle . \label{Expectation}%
\end{equation}
where $\left\vert \Psi_{\mathbf{k}_{c}}\right\rangle $ given by
(\ref{AllState}) is a pure state and the probability for center of mass
momentum $\mathbf{k}_{c}$ is $p_{\mathbf{k}_{c}}$. Normalization is such that
$\left\langle \mathbf{x}_{r}|\mathbf{x}_{r}^{\prime}\right\rangle =\delta
^{3}\left(  \mathbf{x}_{r}-\mathbf{x}_{r}^{\prime}\right)  ,$ $\left\langle
\mathbf{k}_{c}|\mathbf{k}_{c}^{\prime}\right\rangle =\delta_{\mathbf{k}%
_{c},\mathbf{k}_{c}^{\prime}}$ and $\sum_{\mathbf{k}_{c}}p_{\mathbf{k}_{c}%
}=\sum_{\mathbf{k}_{r}}\left\vert c_{\mathbf{k}_{r}}\right\vert ^{2}=$ $\int
d^{3}x_{r}\left\vert \psi\left(  \mathbf{x}_{r},t\right)  \right\vert ^{2}=1$.
The $\Theta$-functions in (\ref{AllState}) limit the volume that the $j^{th}$
photon can occupy to $V=\frac{4}{3}\pi\left(  t-t_{0}\right)  ^{3}$. For
finite volume conservation of momentum, (\ref{kCons}), is approximate, with
uncertainty of order $\pi/\left(  t-t_{0}\right)  $ in each of its components.

\section{Application to experiments}

In this Section, Eq. (\ref{Expectation}) will be applied to Doppler broadening
(PAS experiments) and the arrival time difference between the nuclear photon
and one of the annihilation photons (PAL experiments), and the Irby experiment
will be analyzed.

\subsection{Doppler broadening}

Ref. \cite{Shibuya} reports measurement of the distribution of the Ps center
of mass momentum, so that $\widehat{O}=\left\vert \mathbf{k}_{c}\right\rangle
\left\langle \mathbf{k}_{c}\right\vert $. Substitution in (\ref{Expectation})
gives the probability of center wave vector $\mathbf{k}_{c}$ as%
\begin{equation}
\left\langle \left\vert \mathbf{k}_{c}\right\rangle \left\langle
\mathbf{k}_{c}\right\vert \right\rangle =p_{\mathbf{k}_{c}}. \label{PAS}%
\end{equation}
This experiment was performed using a positron source embedded in biological
tissue, and the Gaussian distribution%
\begin{equation}
p\left(  \mathbf{k}_{c}\right)  =\frac{1}{\left(  \pi\sigma^{2}\right)
^{3/2}}\exp\left(  -\frac{\left\vert \mathbf{k}_{c}\right\vert ^{2}}%
{\sigma^{2}}\right)  \label{c02}%
\end{equation}
with $\sigma=2.4keV=0.005m$ was obtained. The continuous distribution in
related to the discrete probability by $p\left(  \mathbf{k}\right)
=p_{\mathbf{k}_{c}}V/\left(  2\pi\right)  ^{3}.$

If these center of mass momenta were to add coherently, the time uncertainty
for the second photodetection event would be very small. However, the photon
momenta are maximally correlated so, if $\mathbf{x}_{c}$ were to be measured,
(\ref{Expectation}) gives%
\begin{equation}
\left\langle \left\vert \mathbf{x}_{c}\right\rangle \left\langle
\mathbf{x}_{c}\right\vert \right\rangle =\sum_{\mathbf{k}_{c}}p_{\mathbf{k}%
_{c}}\left\vert \left\langle \mathbf{x}_{c}|\mathbf{k}_{c}\right\rangle
\right\vert ^{2}=\frac{1}{V}.
\end{equation}
This implies that the photon center of energy is equally likely to be found
anywhere within the allowed volume, since the only information available about
its position is a consequence of causality and knowledge of the position and
time of positron injection.

\subsection{PAL experiments}

In PAL experiments such as the measurement of positron lifetime in $\alpha
$-$\operatorname{Si}$O$_{2}$ \cite{Saito}, photons are counted at fixed
$\mathbf{x}_{1}$ as a function $t-t_{0}$. It is assumed here that para-Ps
forms as soon as the positron is injected, although in reality the situation
is more complicated than this. To first order in $\beta_{\pm}$ the wave vector
$\mathbf{k}_{r}$ has length $m$ and arbitrary direction. The wave vector
$\mathbf{k}_{c}$ has a definite value and its magnitude is distributed
according to (\ref{c02}). Substitution of $\widehat{O}=\left\vert
\mathbf{x}_{1}\right\rangle \left\langle \mathbf{x}_{1}\right\vert ,$
$\widehat{1}=\int d^{3}x_{2}\left\vert \mathbf{x}_{2}\right\rangle
\left\langle \mathbf{x}_{2}\right\vert $, (\ref{AllState}), and (\ref{Psi_rel}%
) in (\ref{Expectation}) gives%
\begin{align}
\left\langle \left\vert \mathbf{x}_{1}\right\rangle \left\langle
\mathbf{x}_{1}\right\vert \right\rangle  &  =\frac{\Gamma}{4\pi V}\exp\left[
-\Gamma\left(  t-t_{0}\right)  \right] \label{PAL1}\\
&  \times\Theta\left(  t-t_{0}-\left\vert \mathbf{x}_{1}\right\vert \right)
\nonumber\\
&  \times\int d^{3}x_{2}\Theta\left(  t-t_{0}-\left\vert \mathbf{x}%
_{2}\right\vert \right) \nonumber\\
&  \times\left\vert \mathbf{x}_{r}\right\vert ^{-2}\exp\left[  -\Gamma\left(
t-t_{0}-\left\vert \mathbf{x}_{r}\right\vert \right)  \right]  .\nonumber
\end{align}
This is just the trace of the density matrix over the unobserved second
photon. If the $z$-axis is chosen parallel to $\mathbf{k}_{1}$, the
distribution of $\mathbf{k}_{2}$ values is centered at $\theta=\pi$ and the
factor $\exp\left(  \Gamma\left\vert \mathbf{x}_{r}\right\vert \right)  $
selects solid angle $\Omega$ determined by $\Gamma$ and centered about
$\cos\theta=-1$. To first order in $\beta_{\pm}$
\begin{equation}
\left\vert \mathbf{x}_{r}\right\vert =\left\vert \mathbf{x}_{1}\right\vert
+\left\vert \mathbf{x}_{2}\right\vert . \label{x_r}%
\end{equation}
In the limit $\left\vert \mathbf{x}_{1}\right\vert \gg\Gamma^{-1},$ consistent
with our assumption that the pulse is well separated from the source,
$\left\vert \mathbf{x}_{r}\right\vert \approx2\left\vert \mathbf{x}%
_{2}\right\vert $ and the probability density to count a photon at
$\mathbf{x}_{1}$ a time $t-t_{0}$ after positron injection reduces to
\begin{align}
\left\langle \left\vert \mathbf{x}_{1}\right\rangle \left\langle
\mathbf{x}_{1}\right\vert \right\rangle  &  =\frac{\Omega}{16\pi V}\exp\left[
-\Gamma\left(  t-t_{0}-\left\vert \mathbf{x}_{1}\right\vert \right)  \right]
\label{PAL}\\
&  \times\Theta\left(  t-t_{0}-\left\vert \mathbf{x}_{1}\right\vert \right)
\nonumber
\end{align}
where $V=\frac{4}{3}\pi\left(  t-t_{0}\right)  ^{3}$. Thus the rate\ at which
correlated nuclear and annihilation photons are counted decays exponentially.
The coefficient of the exponential reflects our limited knowledge of the
position of the two-photon center of energy.

\subsection{Irby experiment}

In the Irby experiment, illustrated in Fig. 2, photons are emitted by a
source, $S,$ approximately $3mm$ thick. They are detected at the fixed
positions $\mathbf{x}_{1}$ and $\mathbf{x}_{2}$ as a function of $t_{1}-t_{2}$
where $t_{j}$ is the time when a photon is counted at detector $j$. Irby
derived a wave function that generalizes the example considered by Einstein,
Podolsky and Rosen (EPR) by including time dependence and conservation of
energy \cite{IrbyTheory}. He assumed zero center of mass motion so that the
photons have momentum $p$ and $-p$. The relative position, $\mathbf{x}_{r}$,
corresponds to $x_{1}-x_{2}$ and the Fourier amplitude, $c_{\mathbf{k}_{r}},$
given by (\ref{All_c}) corresponds to $f\left(  p\right)  $ in Irby's Eq. (13).

Following EPR\ and Irby \cite{EPR,IrbyTheory} and using (\ref{x_r}) in the
form $\left\vert \mathbf{x}_{r}\right\vert =\left\vert \mathbf{x}%
_{<}\right\vert +\left\vert \mathbf{x}_{>}\right\vert $, the wave function
(\ref{Psi_rel}) can be written as%
\begin{equation}
\psi\left(  \left\vert \mathbf{x}_{r}\right\vert ,t\right)  =\int_{0}^{\infty
}dx\delta\left(  \left\vert \mathbf{x}_{<}\right\vert -x\right)  \psi\left(
\left\vert \mathbf{x}_{>}\right\vert +x,t\right)  \label{delta}%
\end{equation}
where $\delta\left(  \left\vert \mathbf{x}_{<}\right\vert -x\right)  $ is a
positon eigenvector with eigenvalue $x$, $\mathbf{x}_{<}\ $is the position
while $t_{<}$ is the time of the first photodetection event, and
$\mathbf{x}_{>}$ is the position of the second photon. When the first photon
is counted at time $t_{<}$ the wave function collapses to the coefficient of
the $\delta$-function in (\ref{delta}). To ensure propagation at the speed of
light this one-photon exponentially decaying pulse can be written as
\begin{align}
\psi\left(  \left\vert \mathbf{x}_{>}\right\vert ,t\right)   &  =\sqrt
{\frac{\Gamma}{4\pi}}\frac{1}{\left\vert \mathbf{x}_{<}\right\vert +\left\vert
\mathbf{x}_{>}\right\vert }\label{1photon}\\
&  \times\exp\left[  -\frac{1}{2}\left(  iE+\Gamma\right)  \left(  t_{<}%
-t_{0}-\left\vert \mathbf{x}_{<}\right\vert \right)  \right] \nonumber\\
&  \times\exp\left[  -\frac{1}{2}\left(  iE+\Gamma\right)  \left(
t-t_{0}-\left\vert \mathbf{x}_{>}\right\vert \right)  \right]  .\nonumber
\end{align}
Time and distance dependence for the undetected photon is described by the
last exponential, so the probability density is proportional to $\exp\left[
-\Gamma\left(  t-t_{0}-\left\vert \mathbf{x}_{>}\right\vert \right)  \right]
$ or zero. If the second photon is counted at time $t_{>}$, allowing for the
$\mathbf{x}_{c}$ density $V^{-1}$ the probability density for coincident
photodetection is
\begin{equation}
P=\frac{1}{V}\left\vert \psi\left(  \left\vert \mathbf{x}_{r}\right\vert
,\frac{t_{1}+t_{2}}{2}\right)  \right\vert ^{2} \label{P}%
\end{equation}
where $\left\vert \mathbf{x}_{r}\right\vert $ is the detector separation,
$t_{<}+t_{>}=t_{1}+t_{2}$ and $\psi$ is given by (\ref{Psi_rel}).

Essentially the same result is obtained from the second order Glauber
correlation function \cite{Glauber},
\begin{equation}
G^{(2)}\left(  x_{1},x_{2}\right)  =\left\langle E^{(-)}\left(  x_{1}\right)
E^{(-)}\left(  x_{2}\right)  E^{(+)}\left(  x_{2}\right)  E^{(+)}\left(
x_{1}\right)  \right\rangle \label{G2}%
\end{equation}
where $x_{j}=\left(  t_{j},\mathbf{x}_{j}\right)  $. For photodetection at
times $t_{1}$ and $t_{2}$, the positive frequency electric field operators in
$G^{(2)}$ result in a factor
\begin{align}
\exp\left[  -i\left(  \omega_{1}t_{1}+\omega_{2}t_{2}\right)  \right]   &
=\exp\left[  -i\left(  \omega\frac{t_{1}+t_{2}}{2}\right.  \right.
\label{Time}\\
&  \left.  \left.  +\frac{\Delta\omega}{2}\left(  t_{1}-t_{2}\right)  \right)
\right]  \text{.}\nonumber
\end{align}
Since $\sqrt{\omega_{1}\omega_{2}}=m$ is a constant to first order in
$\beta_{\pm}$,
\begin{equation}
G^{(2)}\left(  x_{1},x_{2}\right)  \propto\frac{1}{V}\left\vert \psi\left(
\left\vert \mathbf{x}_{r}\right\vert ,\frac{t_{1}+t_{2}}{2}\right)
\right\vert ^{2} \label{EqualsP}%
\end{equation}
equal to $P$ given by (\ref{P}).

The probability density $P$ is proportional to $\exp\left[  -\Gamma\left(
t_{1}+t_{2}-2t_{0}\right)  \right]  $, but Irby measured the distribution of
$t_{1}-t_{2}$, and neither (\ref{P}) nor the absolute square or Irby's wave
function in \cite{IrbyExpt} gives their probabilities directly. The resolution
to this problem lies in averaging over the positron injection time,\emph{
}$t_{0}$,\ that is not measured but must be earlier than both $\tau_{1}$ and
$\tau_{2}$. If it is assumed that positrons are injected at a constant rate
$r=1/T$ , substitution of (\ref{Psi_rel}) in (\ref{P}) gives%
\begin{align}
P &  =\frac{r\Gamma}{4\pi\left\vert \mathbf{x}_{r}\right\vert ^{2}V}%
\int_{-T/2}^{T/2}dt_{0}\label{Irby}\\
&  \times\exp\left[  -\Gamma\left(  t_{1}+t_{2}-2t_{0}-\left\vert
\mathbf{x}_{r}\right\vert \right)  \right]  \nonumber\\
&  \times\Theta\left(  \tau_{1}-t_{0}\right)  \Theta\left(  \tau_{2}%
-t_{0}\right)  .\nonumber
\end{align}%
\begin{figure}
[ptb]
\begin{center}
\includegraphics[
height=0.8112in,
width=3.2491in
]%
{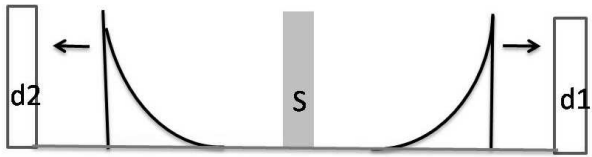}%
\caption{Irby experiment. A positron is created in the source, S, and the time
difference between annihilation photons arriving at detectors d1 and d2 is
measured.}%
\label{FIG. 1}%
\end{center}
\end{figure}
The integral (\ref{Irby}) is evaluated as $I3$ in Appendix C with the upper
limit of the $t_{0}$ integral is taken to be the earlier photon emission time.
The result is%
\begin{equation}
P=\frac{r}{8\pi\left\vert \mathbf{x}_{r}\right\vert ^{2}V}\exp\left(
-\Gamma\left\vert \tau_{1}-\tau_{2}\right\vert \right)  ,\label{Irby1}%
\end{equation}
where $\tau_{j}=t_{j}-\left\vert \mathbf{x}_{j}\right\vert $.

Irby fit his date to a Lorentzian curve while, according to (\ref{Irby1}), the
experimental picosecond timing analyzer (PTA) spectrum in Figs. 4 and 5 of
Ref. \cite{IrbyExpt} is a double exponential. This discrepancy is addressed in
Fig. 3 that shows a comparison of a double exponential to a Lorentzian and a
Gaussian. The double exponential gives the sharp peaks observed by Irby while
behaving like the Lorentzian that he used in his fits in the tails. The
Gaussian has an appreciably different shape and does not fit the data as noted
by Irby. Eq. (\ref{Irby1}) derived here should give an improved description of
the experimental results.
\begin{figure}
[ptb]
\begin{center}
\includegraphics[
height=1.6786in,
width=3.0761in
]%
{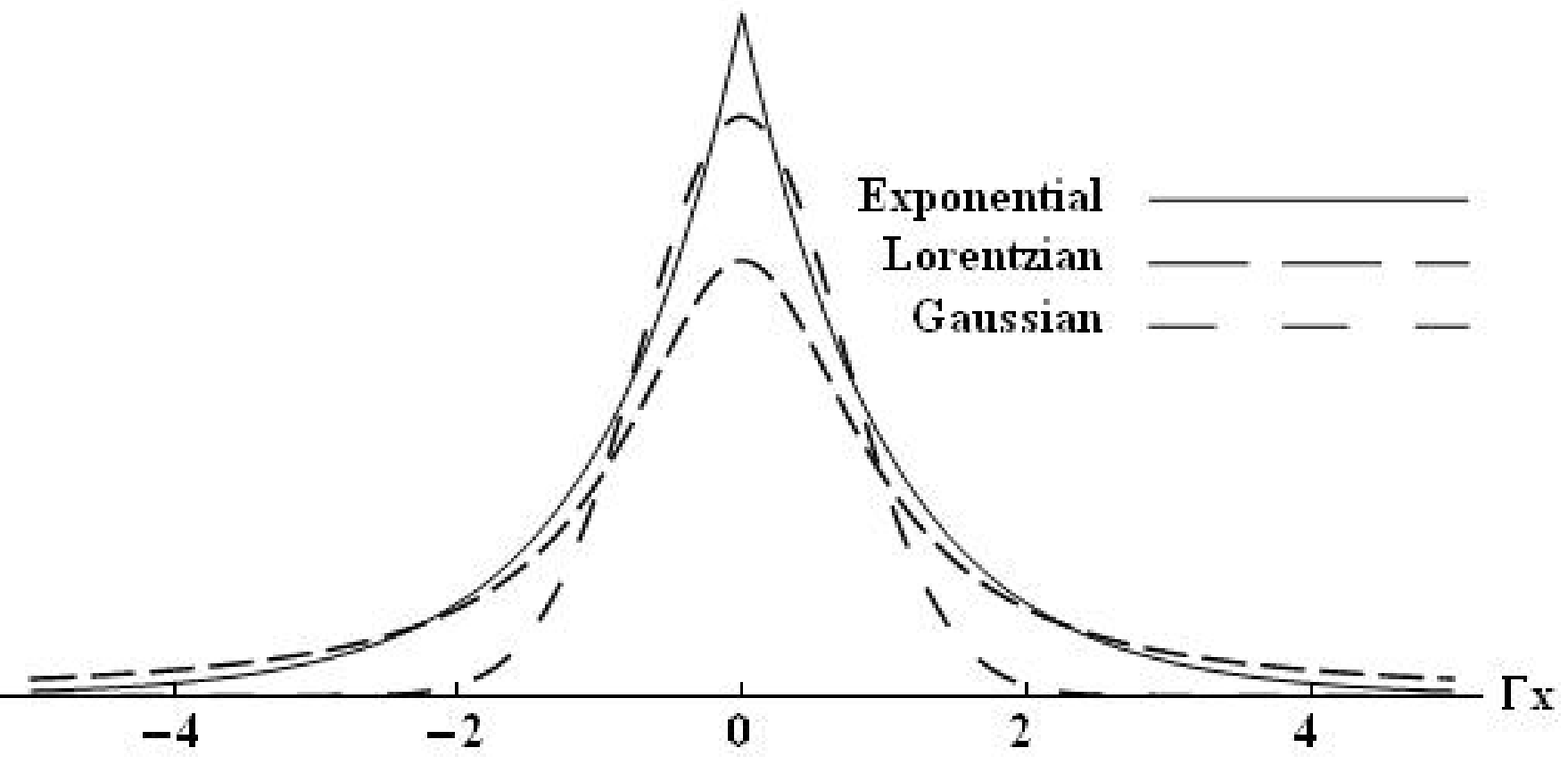}%
\caption{Comparison of exponential of the absolute value, $\left(
2\Gamma\right)  ^{-1}\exp\left(  -\Gamma\left\vert x\right\vert \right)  $
with a Lorentzian, $\left(  \pi\Gamma^{2}\right)  ^{-1}\left(  x^{2}%
+\Gamma^{2}\right)  ^{-1},$ and a Gaussian, $\left(  \Gamma\pi\right)
^{-3/2}\exp\left(  -\Gamma^{2}x^{2}\right)  $.}%
\label{FIG. 3}%
\end{center}
\end{figure}

\section{Conclusion}

This paragraph describes the details of the present calculation in relation to
the previous theoretical work: In Refs. \cite{EPR} and \cite{IrbyTheory} the
center of energy momentum is set equal to zero, the wave function is given as
a function of the relative coordinates, and the time during which the photons
interact, here $t_{0}$ to $t_{0}+\Gamma^{-1}$, is assumed to be known. In the
present calculation the momentum of the center of energy has a wide range of
definite values consistent with the PAS experiments, and the positron
injection time is unknown. In \cite{EPR} all relative momenta are given equal
weight. Since the time when the particles interact is known, when one of the
counterpropagating photons is detected the position of the second photon is
determined exactly and nonlocally by collapse of the wave function. Here and
in \cite{IrbyTheory} the relative momenta, $p=\left\vert \mathbf{k}%
_{r}\right\vert $, are restricted by a function $f\left(  p\right)  $ which we
find here is a Lorentzian with center at $\left\vert \mathbf{k}_{r}\right\vert
=m$ and FWHM $2\Gamma$, resulting in exponential decay in space-time.

Irby attributed the unexpectedly wide range of annihilation photon PTA
detection time differences that he observed to maximally restricted photon
momenta, leading to the elimination of nonlocality in the conjugate position
observables \cite{IrbyTheory}. Here the pure states have definite center of
energy momentum and Ps decay is described in terms of the relative
coordinates. After averaging over the unobserved positron injection time, the
annihilation photon coincidence rate was found to be proportional to
$\exp\left(  -\Gamma\left\vert \tau_{1}-\tau_{2}\right\vert \right)  $ where
$\tau_{j}$ is the photon emission time. This supports Irby's observation
\cite{IrbyExpt} that annihilation photon pulse width is limited by the Ps
lifetime. Only the peak of the double exponential function is determined by
the position of the positron source. This is counter to expectations,
and\ should be taken into account in TOF PET imaging.

Annihilation photons have played a significant role in the development of our
understanding of quantum correlations. Their polarization correlations were
considered, and discarded, as a candidate for the first experimentally
realizable test of Bell's theorem \cite{Bell}. EPR used position correlations
of a pair of counterpropagating particles as their primary example of nonlocal
collapse \cite{EPR}. Irby performed a direct measurement of annihilation
photon space-time correlations and concluded that their nonlocality is erased
by maximal restriction of their momenta. Here we find that their momenta are
maximally correlated because their center of energy momentum has a well
defined value. Position entanglement is ascribed to the relative coordinates,
augmented by causality. The observed $123\ ps$ pulse width is attributed to
uncertainty in the time of photon pair creation due to Ps annihilation.

\begin{acknowledgments}
The authors thank the Natural Sciences and Engineering Research Council for
financial support.
\end{acknowledgments}

\appendix{}

\section{Relative normalization}

Normalization requires evaluation of%
\[
I1=\sum_{\mathbf{k}}\frac{1}{\left(  \omega-E\right)  ^{2}+\Gamma^{2}}%
=\frac{V}{2\pi^{2}}\int_{0}^{\infty}dkk^{2}\frac{1}{\left(  \omega-E\right)
^{2}+\Gamma^{2}}.
\]
Since $\omega_{c}\approx2\left\vert \mathbf{k}_{r}\right\vert $ according to
(\ref{mag_k}), we want $\omega=2k$. Making a change of variables to
$\eta=2k-E$ with limits $-\infty$ to $\infty$ and selecting a contour that
encloses the pole at $\eta=-i\Gamma$ with $\Gamma\ll E$ gives%
\[
I1=\frac{V}{2\pi^{2}}\left(  \frac{E}{2}\right)  ^{2}\frac{2\pi i}{4i\Gamma
}=\frac{VE^{2}}{16\pi\Gamma}.
\]

\section{Relative k- to x-space integral}

To evaluate (\ref{Relative_x}) we need%
\begin{align*}
I2  &  =\sqrt{\frac{16\pi\Gamma}{V^{2}E^{2}}}\int d^{3}k\frac{\exp\left(
i\mathbf{k}\cdot\mathbf{r}-i\omega t\right)  }{\omega-E+i\Gamma}\\
&  =\sqrt{\frac{16\pi\Gamma}{E^{2}}}\frac{2\pi}{ir}\int_{0}^{\infty}%
dkk\frac{\exp\left(  ikr\right)  -\exp\left(  -ikr\right)  }{2k-E+i\Gamma}\\
&  \times\exp\left(  -i2kt\right)  .\\
&  =\sqrt{\frac{\Gamma}{4\pi}}\frac{1\ }{r}\left\{  \exp\left[  -\left(
iE+\Gamma/2\right)  \left(  t-\frac{1}{2}r\right)  \right]  \right. \\
&  \left.  -\exp\left[  -\left(  iE+\Gamma\right)  \left(  t+\frac{1}%
{2}r\right)  \right]  \right\}
\end{align*}

\section{Irby experiment $t_{0}$-integral}

We need%
\begin{align*}
I3  &  =\int_{-T/2}^{T/2}dt_{0}\exp\left[  -\Gamma\left(  2t_{c}%
-2t_{0}-\left\vert \mathbf{x}_{r}\right\vert \right)  \right] \\
&  \times\Theta\left(  \tau_{1}-t_{0}\right)  \Theta\left(  \tau_{2}%
-t_{0}\right)
\end{align*}
If $T\gg\Gamma^{-1}$ the limits can be extended to $\pm\infty$ and the
$\Theta$-functions imply that%
\begin{align*}
I3  &  =\exp\left[  -\Gamma\left(  2t_{c}-\left\vert \mathbf{x}_{r}\right\vert
\right)  \right]  \int_{-\infty}^{\tau_{<}}dt_{0}\exp\left[  2\Gamma
t_{0}\right] \\
&  =\left(  2\Gamma\right)  ^{-1}\exp\left[  -\Gamma\left(  2t_{c}-2\tau
_{<}-\left\vert \mathbf{x}_{r}\right\vert \right)  \right]
\end{align*}
where $\tau_{>}$ $(\tau_{<})$ is the larger (smaller) of $\tau_{1}$ and
$\tau_{2}.$ Since according to (\ref{coord}) and (\ref{tau})
\begin{align*}
2t_{c}-2\tau_{<}  &  =t_{>}+t_{<}-2t_{<}+2\left\vert \mathbf{x}_{<}\right\vert
\\
&  =t_{>}-t_{<}+2\left\vert \mathbf{x}_{<}\right\vert ,
\end{align*}%
\[
I3=\left(  2\Gamma\right)  ^{-1}\exp\left[  -\Gamma\left(  t_{>}%
-t_{<}-\left\vert \mathbf{x}_{r}\right\vert +2\left\vert \mathbf{x}%
_{<}\right\vert \right)  \right]  .
\]
Eq. (\ref{x_r}) gives $\left\vert \mathbf{x}_{r}\right\vert =\left\vert
\mathbf{x}_{>}\right\vert +\left\vert \mathbf{x}_{<}\right\vert $, that is the
distance between the detectors equals the sum of the source-detector
distances, so that%
\begin{align*}
I3  &  =\left(  2\Gamma\right)  ^{-1}\exp\left[  -\Gamma\left(  t_{>}%
-t_{<}-\mathbf{x}_{>}+\left\vert \mathbf{x}_{<}\right\vert \right)  \right] \\
&  =\left(  2\Gamma\right)  ^{-1}\exp\left[  -\Gamma\left(  \tau_{>}-\tau
_{<}\right)  \right]  .
\end{align*}

\end{document}